\documentclass[a4paper,11pt]{article}
\usepackage{jinstpub} 
\usepackage{amsfonts}
\usepackage{amsmath}
\usepackage{amsthm}
\usepackage{fancyhdr}
\usepackage{graphicx}
\usepackage{color}
\usepackage{colortbl}
\usepackage{quoting}
\usepackage{multirow}
\usepackage{cancel}
\usepackage{soul}
\usepackage{gensymb}
\usepackage{xr}
\usepackage{siunitx}
\usepackage{lscape}
\usepackage[acronym]{glossaries}
\usepackage{caption}
\usepackage{subcaption}
\usepackage{cleveref}

\usepackage{lineno}

\title{Characterization of the performances of commercial plastic scintillators in cryogenic environments}

\author[a]{M. Biassoni,}
\author[b]{A. Caminata,}
\author[b]{S. Caprioli,}
\author[b]{A. Celentano,}
\author[b]{S. Davini,}
\author[b,c,1]{A. Marini,}
\author[b]{G. Sobrero}
\note{Corresponding author.}

\affiliation[a]{Istituto Nazionale di Fisica Nucleare, sezione di Milano Bicocca, Piazza della Scienza, 3 - I, Milano, Italy.}
\affiliation[b]{Istituto Nazionale di Fisica Nucleare, sezione di Genova, via Dodecaneso 33, Genova, Italy.}
\affiliation[c]{Università degli Studi di Genova, Dipartimento di Chimica e Chimica Industriale, via Dodecaneso 31, Genova, Italy.}

\emailAdd{anna.marini@ge.infn.it}

 \abstract{Plastic scintillators have become increasingly important in particle physics for time-of-flight and calorimetry measurements. Their light yield and the possibility of customizing their geometry make them also attractive for the construction of active vetoes in rare event physics experiments. For this purpose, some commercial plastic scintillators (purchased from Eljen Technology) were tested in cryogenic environments (liquid nitrogen and liquid helium). Their relative light yield was estimated by comparing the data acquired at room temperature with those acquired at cryogenic temperatures. Finally, estimates of the variation of the light yield at cryogenic temperatures were obtained.}

\arxivnumber{2303.03008}

\begin{document}
\maketitle
\flushbottom

\section{Introduction}
The use of plastic scintillators is a well-established technique for particle detection. Thanks to the light yield (LY) of about $10^4$ photons$/$MeV in case of energy released by an electron~\cite{Plastic_scint} and the possibility to manufacture them in different shapes, they could be used as particle taggers in different scenarios. 
The main use of these devices is for time-of-flight apparata~\cite{Berdugo:2022zzo}~\cite{ToF} and calorimetric measurements (like the ATLAS sampling calorimeter~\cite{ALEKSA2013442}).
Further applications for these devices are, for example, their use as scintillating optical fibers, as is the case for the neutrinoless double beta decay experiment GERDA~\cite{Gerda}, or as a veto system (as it is foreseen in LEGEND-200~\cite{Legend_design}, which led to the creation of devices with an attenuation length of about 6 cm, which is sufficiently large for this application).
To realize larger apparatuses it is advisable to modulate the attenuation length accordingly. \\
Their use in rare event experiments derives from the possibility of realizing devices with a high degree of radiopurity by selecting radiopure starting materials (which are in the form of liquids) and carrying out the polymerization process (which requires the addition of some catalysts and possibly an increase in pressure and/or temperature) in controlled environments, using clean containers.
At the moment, the devices developed for the SuperNEMO experiment~\cite{NEMO} are the state of the art in terms of radiopure plastic scintillators. \\
Another important feature that makes these scintillators suitable for a large range of applications, is their high hydrogen content, thanks to which they can act as neutron moderators for dark matter search noble liquids experiments. \\
In the case of a direct search for dark matter, it is also essential to tag the neutrons, in addition to capturing them, therefore it is possible to think of an application of these devices as an active veto. 
This work aims to test commercial plastic scintillators at cryogenic temperatures (the tests were conducted in LN$_2$, where the devices reached a minimum temperature of 79 K, and LHe, where the minimum temperature was around 20 K), both for their resistance to low temperatures and for their effective LY. As far as resistance to low temperatures is concerned, some studies on the use of plastic materials in the cold as a structural material are reported in the literature~\cite{refId0}. To date, there is no information in the literature on the performance of plastic scintillators under these thermal conditions. Therefore, it was considered of scientific interest to characterize them in a cryogenic environment, to verify if this can be the starting point for the realization of radiopure devices suitable for rare-events physics experiments. Six different scintillators, purchased from Eljen Technology~\cite{Eljen}, were tested. Their main physical properties are reported in tab. \ref{tab:PESCE_Eljen_scintillators}.
\begin{table}[!h]
    \centering
    \begin{tabular}{ccccc}
    \hline
    Scintillator & \begin{tabular}[c]{@{}c@{}}Scintillating \\ molecule \\{[\%]}~\end{tabular} & \begin{tabular}[c]{@{}c@{}}Light \\yield\\{[}$\gamma/$1 MeV e$^-$]\end{tabular} & \begin{tabular}[c]{@{}c@{}}Light attenuation \\length \\{[}cm]\end{tabular} & \begin{tabular}[c]{@{}c@{}}Emission\\wavelenght\\{[}nm]\end{tabular} \\ \hline \hline
        EJ-200 & 64 & 10000 & 380 & 425 \\
        EJ-208 & 60 & 9200 & 435 & 408 \\
        EJ-230 & 64 & 9700 & 120 & 391 \\
        EJ-240 & 41 & 6300 & 240 & 430 \\
        EJ-244 & 56 & 8600 & 270 & 434 \\
        EJ-248 & 60 & 9200 & 250 & 425 \\ \hline 
        \end{tabular}
    \caption{Main characteristics of the studied commercial organic scintillators~\cite{Eljen}.}
    \label{tab:PESCE_Eljen_scintillators}
\end{table}

All the scintillators under study had been characterized by the vendor in a thermal range between -20\,\celsius\, and 60\,\celsius\, and it was stated that there were no changes in the light output from -60\,\celsius\ to 20\,\celsius.

\section{Experimental set-up}
\begin{figure}[!h]
    \centering
    \includegraphics[height=3.5 in]{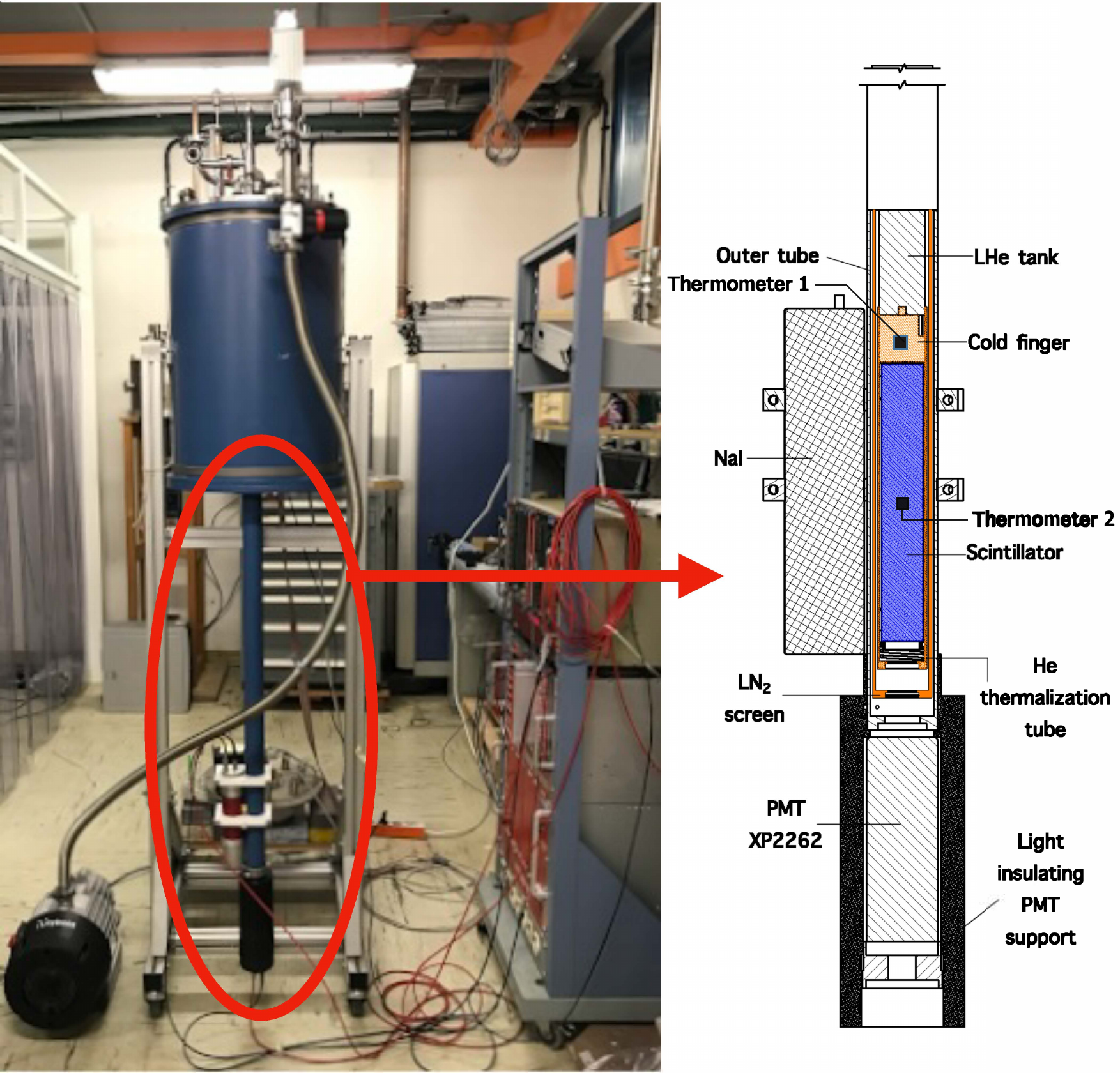}
    \caption{Schematic of the experimental setup. The white holders attached to the blue tube act as housing both for the inorganic scintillator and the PMT coupled to it and for the $\gamma$s source, that is the $^{60}$Co, in order to keep it in a fixed position easily. The cold finger is a copper tube that is in direct contact both with the plastic scintillator and with the cryogenic bath in the cryostat.}
    \label{fig:Schema_tubo}
\end{figure}

\begin{figure}[!h]
\centering
\begin{subfigure}{0.6\textwidth}
    \includegraphics[width=\textwidth]{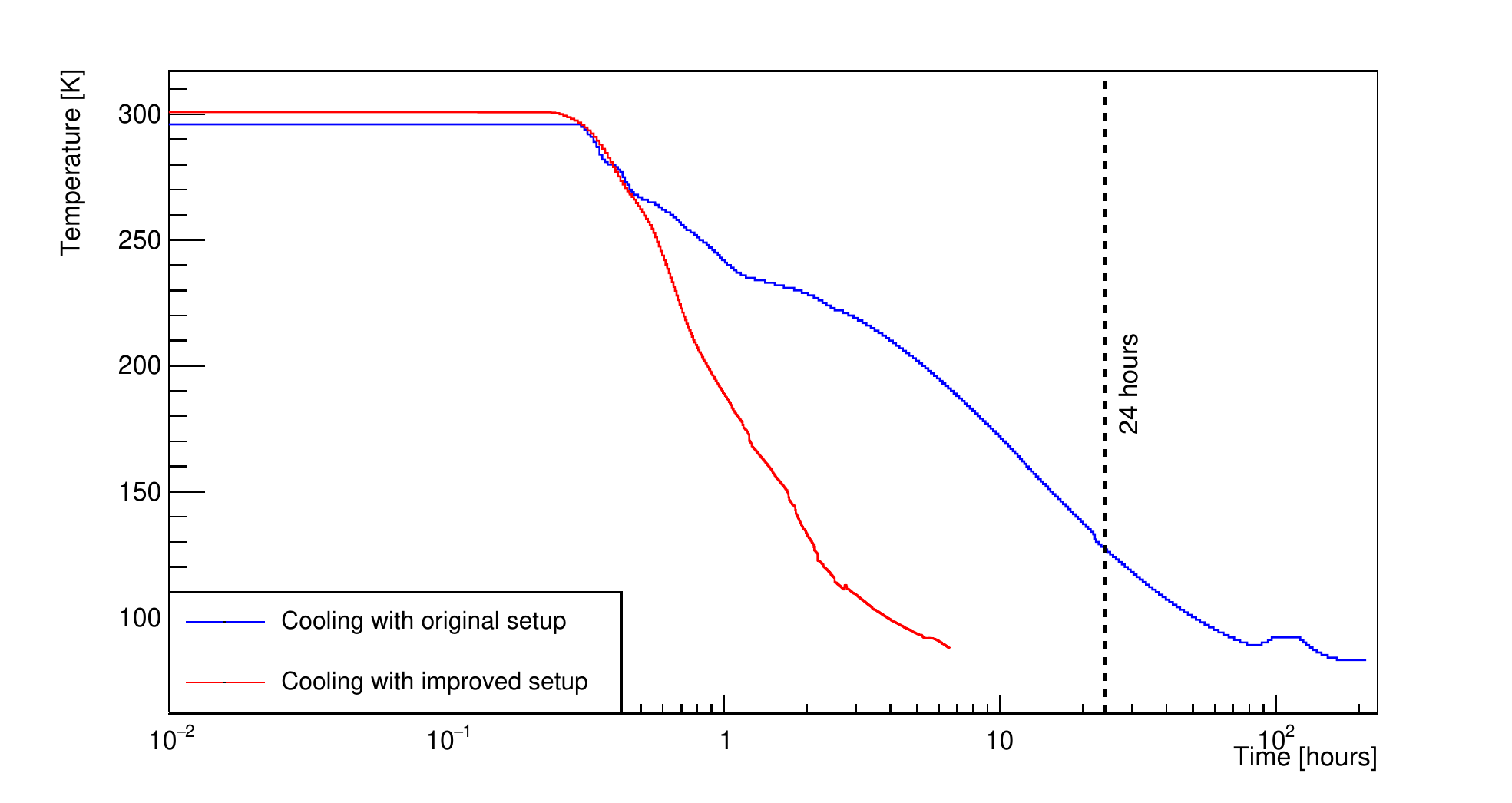}
    \caption{Difference in cooling time between different thermal links. The black dashed line is positioned at a time equal to 24 hours since the cooling started.}
    \label{fig:PESCE_cooling_plot}
\end{subfigure}
\hfill
\begin{subfigure}{0.35\textwidth}
    \includegraphics[width=\textwidth]{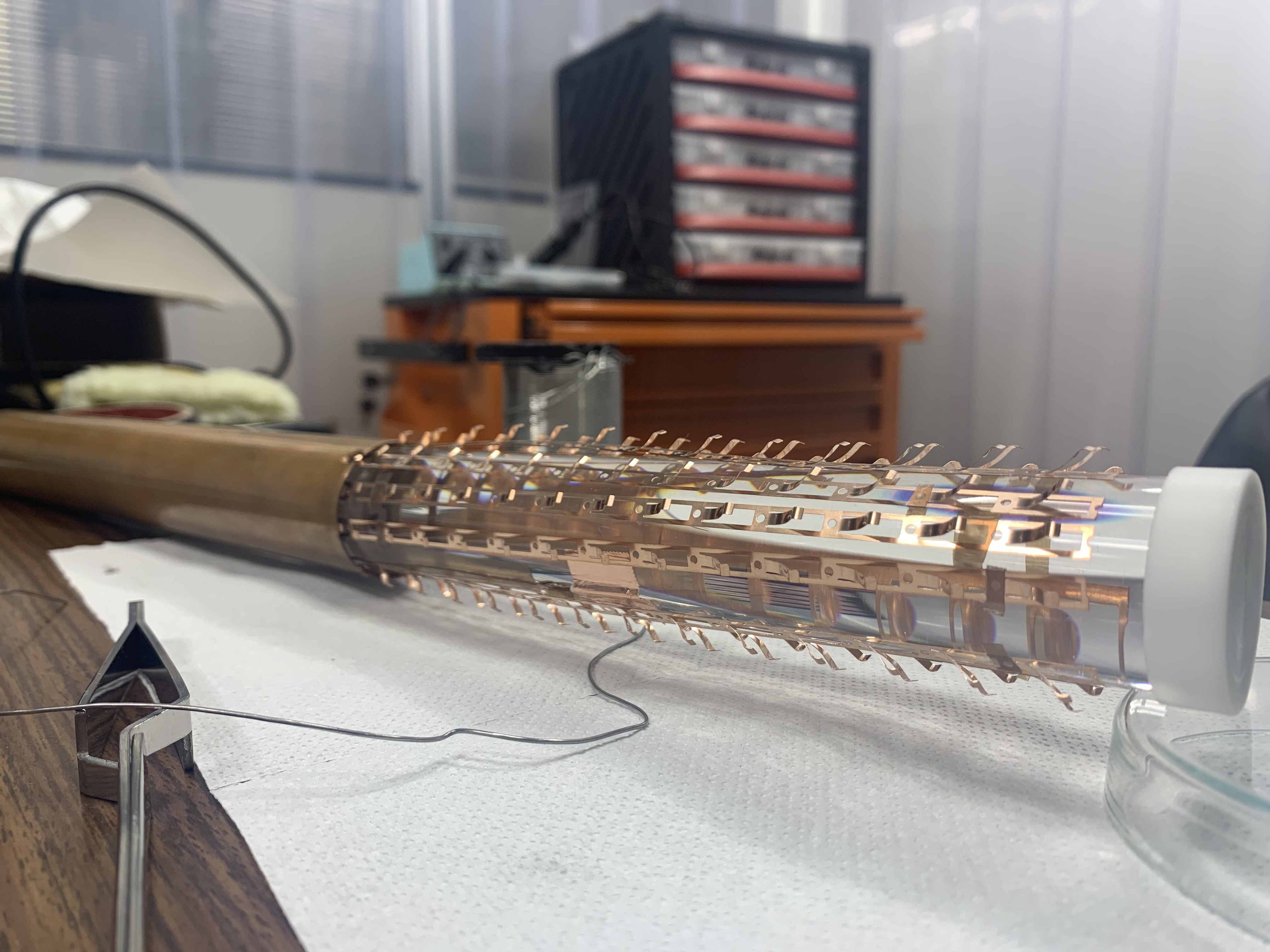}
    \caption{Placing of copper-beryllium strings along the scintillator profile to optimize the thermal coupling from the cold finger to the scintillator.}
    \label{fig:PESCE_lamelle}
\end{subfigure}
\caption{Different cooling set-up for the data taking. In fig.~\ref{fig:PESCE_cooling_plot}, we can see the consistent difference in the cooling time between the two different set-ups. In fig.~\ref{fig:PESCE_lamelle}, instead, an example of the typical configuration of the lamellae on the plastic scintillator is reported.}
\label{fig:PESCE_Cooling_setup}
\end{figure}

The tests were conducted in a double-wall cryostat by Oxford Instruments, equipped with a copper rod (cold finger) to ensure thermal contact between the scintillator and the cryogenic liquid bath. The vacuum has been stably maintained in the cryostat by means of a rotary pump Leybold Scrollvac 18 plus, capable of bringing the pressure down to 10$^{-2}$\,mbar. The vacuum level was measured thanks to a Leybold Ionivac ITR200S meter read by the Leybold Graphix controller.  
The experimental set-up (shown in fig.~\ref{fig:Schema_tubo}) involved the use of two scintillators: the scintillator under exam and an external NaI crystal doped with Tl (by Scionix, 51B51/2M-E12"), coupled with a photomultiplier tube, used for triggering.
The plastic scintillator was coupled to a photomultiplier tube (PMT, Philips XP2262B), and both the PMTs were connected to an 8-bit and 500 MS/s digitizer (CAEN v1731). 
With this experimental setup, it was not possible to perform an optimal optical coupling between the scintillator and the PMT, since the first one was placed inside the cryostat. For this purpose, there are two optical windows on the end of the cryostat tube, which however guarantee an imperfect optical coupling and, consequently, a poor collection of light.
We installed Pt-100 sensors and DT-670 diodes to monitor the temperature of the cold finger and the surface of the plastic scintillator under test. Pt-100 sensors were used during the commissioning phase at LN$_2$ and we upgraded to DT-670 diodes for the measurement. The temperature sensors were read and recorded through a LakeShore 218 temperature monitor or a Cryocon 18i in a different configuration of the slow monitoring system. To characterize the light yield of the scintillator, we used a 1.26\,kBq $^{60}$Co-based source, which emits two simultaneous $\gamma$-rays, with energies of 1.17\,MeV and 1.33\,MeV, respectively~\cite{60Co}.
The acquisition was performed by triggering on the NaI signal: if the signal detected by the NaI reached a certain amplitude threshold, an acquisition window of 8 $\mu$s was opened for the plastic scintillator, long enough to identify the signal, the background noise level, and possible accidental coincidences. \\
The signals of the two PMTs were then recorded as waveforms by a customized data acquisition system developed using LabView software~\cite{Labview}. The data files have been reconstructed and analyzed offline with dedicated software.
The data reconstruction software is structured in various modules run sequentially: the first module reads the data file and reconstructs the waveforms in physical units (amplitude in Volts, timescale in nanoseconds); the second module estimates the mean and standard deviation of the pedestal from the first 250\,ns of the acquisition; finally, the third module computes the maximum amplitude and the integral of the waveform in specific regions of interest: a region for pedestal estimation, 250\,ns wide, a region for the scintillator signal estimation, 250\,ns wide, starting 100\,ns after the pedestal region, and a region for the NaI trigger signal, 400\,ns wide, after the scintillation region.
The scintillation region is large enough to include the whole signal; the same length is also used for the pedestal region, to simplify the comparison of the two regions in the analysis.
All the reconstructed information of the event is then stored in a file based on the CERN ROOT~\cite{ROOT} toolkit, which contains both the information of the event (e.g. pedestal, amplitude, charge) and the slow monitoring parameters (e.g. temperature). \\
The time needed for cooling with cryogenic liquids strictly depends on the goodness of the thermal coupling between the scintillator and the cold finger. In the first tests we performed, the cold finger was placed in contact with a PTFE layer, but this led to unsatisfactory results, as the scintillator reached temperatures close to those of liquid nitrogen in a several-day time frame
(see blue curve in fig.~\ref{fig:PESCE_cooling_plot}). 
After that, the thermal coupling was optimized by thinning the PTFE layer and placing some copper-beryllium strings on the scintillator surface (see fig. \ref{fig:PESCE_lamelle}), which improved the thermalization. As seen from the red curve in fig.~\ref{fig:PESCE_cooling_plot}, the cooling time has been significantly reduced as a result of this upgrade.
As regards the thermal stability during the acquisitions, the cryostat, together with the vacuum system, have allowed us to reach a satisfactory level of insulation. The positioning of the thermometers allowed us to acquire the temperature values during the runs. Taking as an example one of the longest LN$_2$ runs (about 100 hours), the average temperature was (79.19 $\pm$ 0.17) K.

\section{Data taking}
\begin{figure}[!h]
\centering
\begin{subfigure}{0.45\textwidth}
    \includegraphics[width=\textwidth]{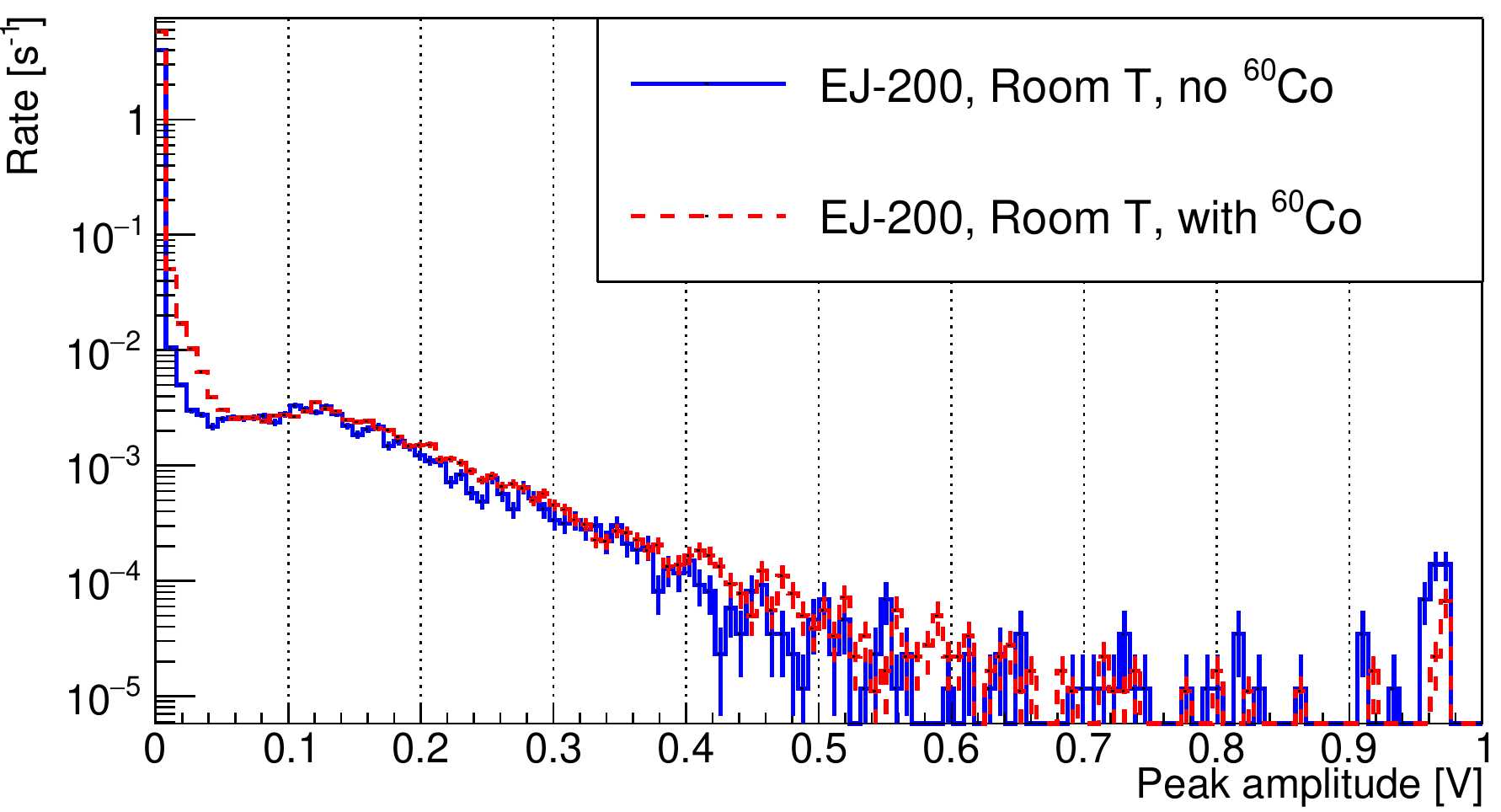}
    \caption{The presence of the $^{60}$Co source is visible at low energies (see fig.~\ref{fig:EJ-200_ZOOM}). At high energies, the contribution of muons can be seen in both curves.}
    \label{fig:EJ-200}
\end{subfigure}
\hfill
\begin{subfigure}{0.45\textwidth}
    \includegraphics[width=\textwidth]{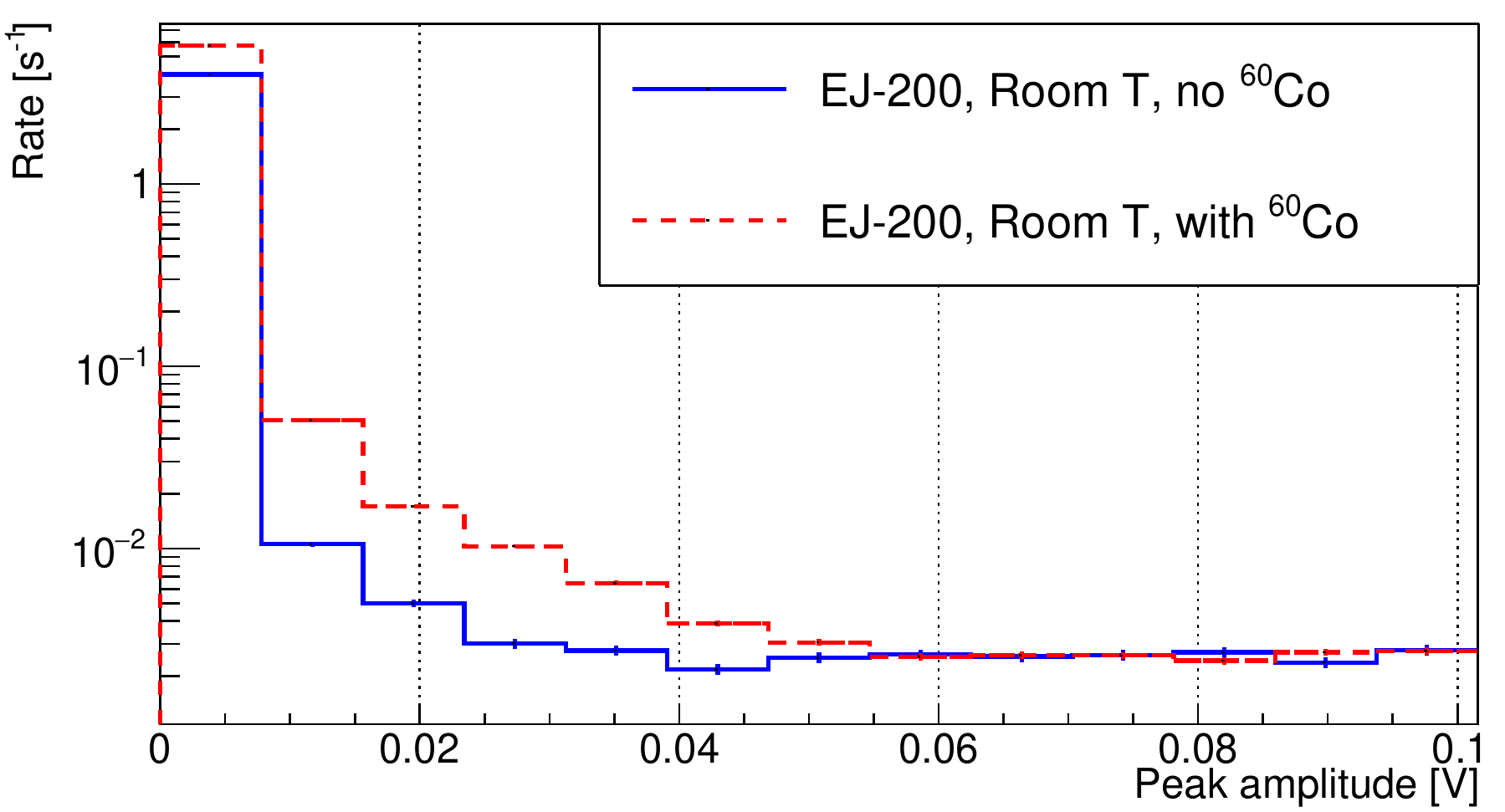}
    \caption{Zoom of fig.~\ref{fig:EJ-200}: evidence of the effect of the $^{60}$Co source at low energies, visible as an excess of events between 0.01-0.06 V for the red curve.}
    \label{fig:EJ-200_ZOOM}
\end{subfigure}
\caption{Typical shape of the energy plot. The blue curve represents a spectrum acquired at room temperature, without the $^{60}$Co source, while the red curve represents a room temperature spectrum acquired in presence of the $^{60}$Co source.}
\end{figure}

The light yield variation of each scintillator sample from ambient to cryogenic temperature has been studied by comparing energy spectra 
acquired during different runs.
A typical example is reported in fig.~\ref{fig:EJ-200}, where we can see the spectra of the EJ-200 sample at room temperature, with and without the radioactive source, whose effect at low energies is enlighted in fig.~\ref{fig:EJ-200_ZOOM}. Each spectrum is normalized to the total duration of the respective acquisition run.
For reference, the single photoelectron region lies around 12\,mV.
Each spectrum is typically made up of three parts:
\begin{itemize}
    \item a steep drop in the (0.00-0.01)\,V, due to electronic noise and accidental coincidences with dark counts.
    \item a shoulder in the region (0.01-0.06)\,V produced by the $^{60}$Co source (absent in the spectrum with no source).
    \item a shoulder in the (0.06-0.5)\,V range, due to cosmic muons.
\end{itemize}

\begin{figure}[!h]
    \centering
    \includegraphics[width=0.8\textwidth]{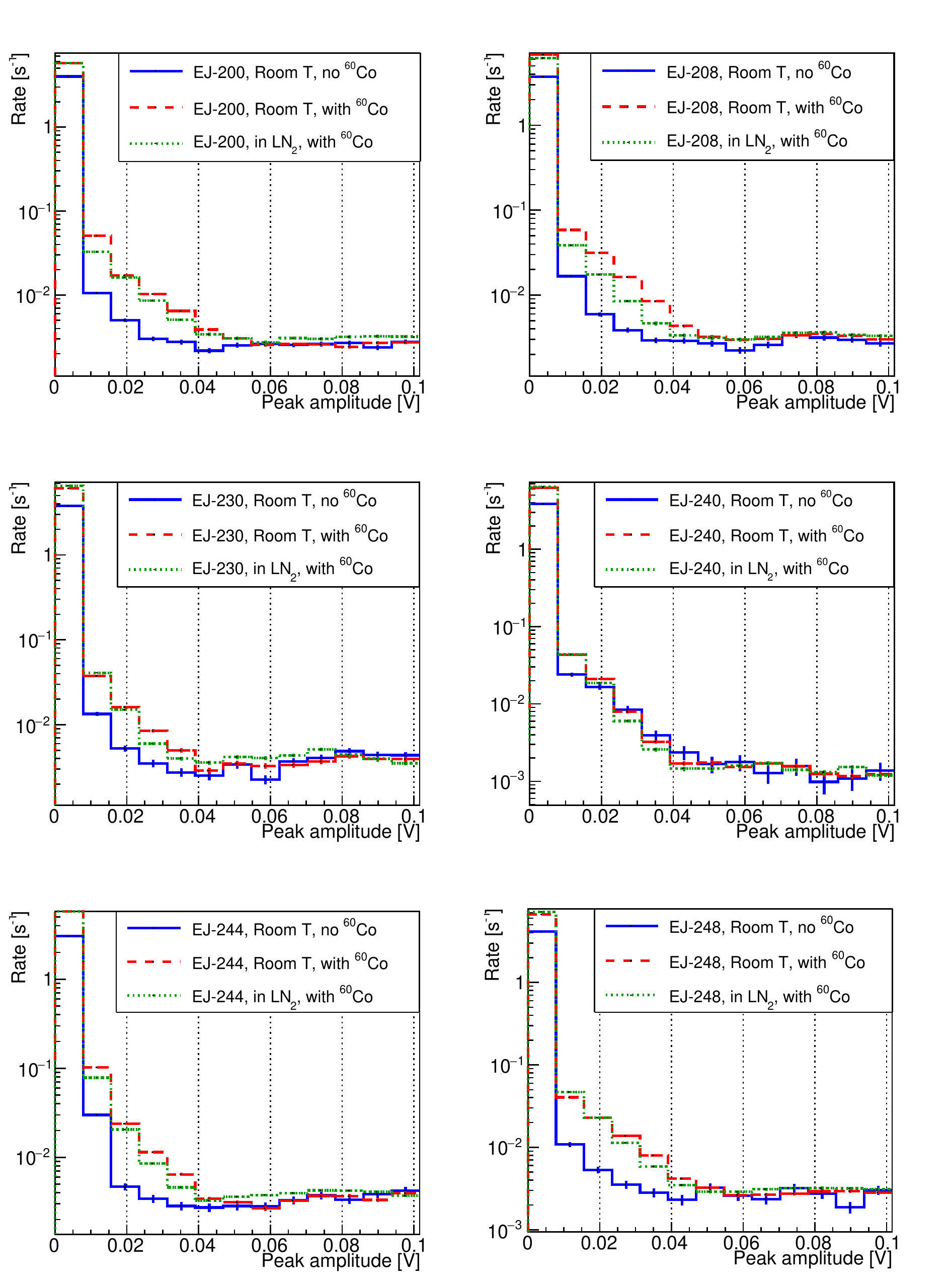}
    \caption{Comparison of the efficiency of all scintillators at room temperature and at temperatures close to those of nitrogen liquefaction. The blue curve shows, for all devices, a run performed in the absence of the $\gamma$-ray source at room temperature. The dashed red curve represents a run acquired at room temperature in the presence of the $\gamma$-ray source. Finally, the green dotted curve shows a run in the presence of the $\gamma$-ray source acquired in a liquid nitrogen environment.}
    \label{fig:LN2}
\end{figure}

Cooling tests in LN$_2$ were made for all the scintillators in tab.~\ref{tab:PESCE_Eljen_scintillators}. The effect of the temperature on the rate of scintillation events was different for each device, as  shown in fig.~\ref{fig:LN2}. This was expected since they carry different concentrations of the scintillating molecule.

\begin{figure}
    \centering
    \includegraphics[width=0.5\textwidth]{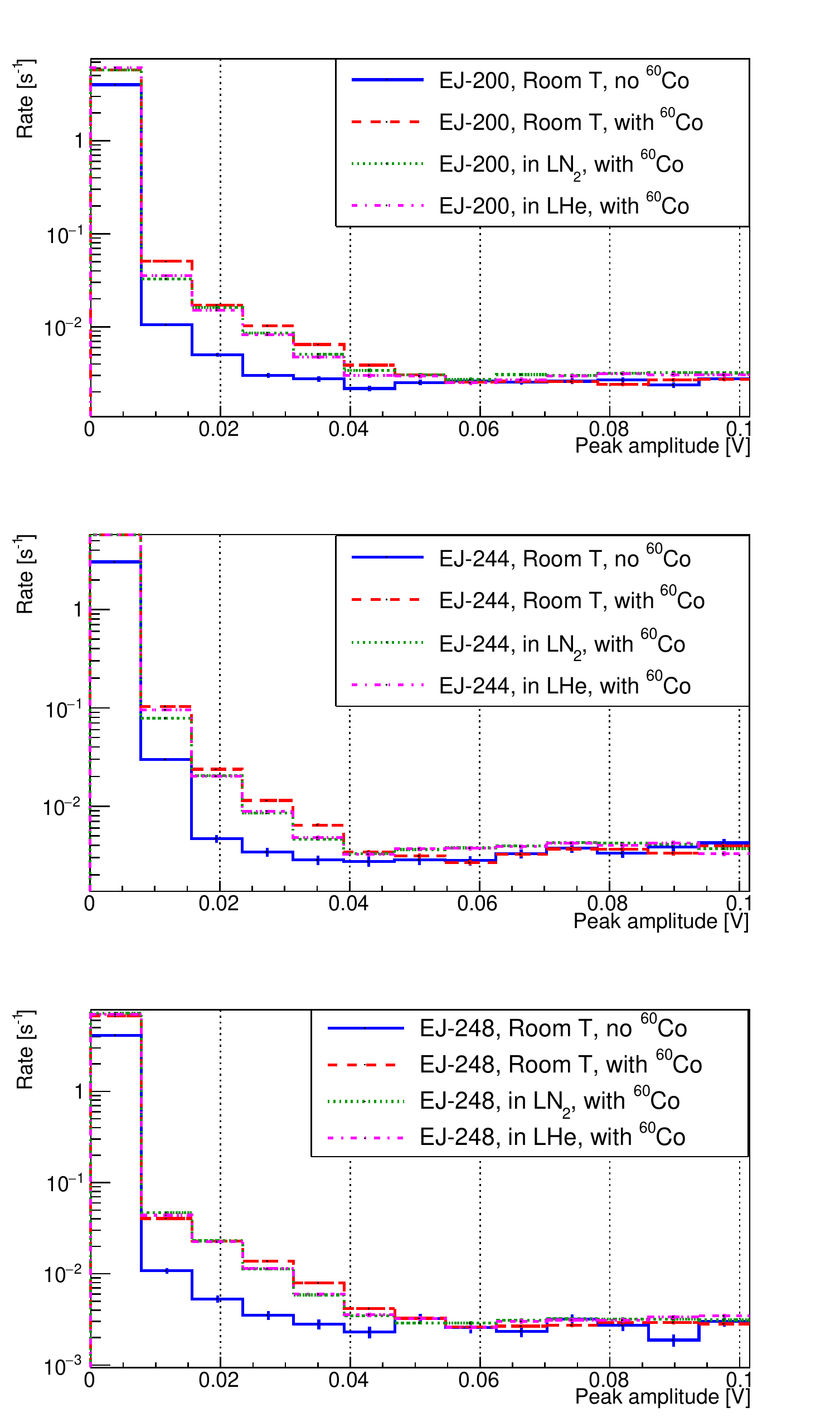}
    \caption{Overview of the performances of the scintillators of EJ-200, E-J244, and EJ-248 at room temperature, in a liquid nitrogen environment, and in a liquid helium environment (dashed and dotted magenta curves).}
    \label{fig:LHe}
\end{figure}

In view of the results obtained cooling the scintillators in LN$_2$ (whose results are reported in section~\ref{Analysis}), it was decided to characterize in LHe the three scintillators that most of all showed a good LY and a high light attenuation length at the same time, since the choice of attenuation length is a relevant parameter in the construction of an apparatus for the physics of rare events. The chosen scintillators for this purpose are EJ-200, EJ-244, and EJ-248. 
The plots are illustrated in fig.~\ref{fig:LHe}.
Even at such low temperatures, no drastic decreases in efficiency are noted for these scintillators, since the curves relating to the acquisition at LHe do not seem to show variations compared to the data that were taken at LN$_2$ (dashed red and dotted green curves).

\section{Data analysis}\label{Analysis}
\begin{figure}[!h]
    \centering
    \includegraphics[width=\textwidth]{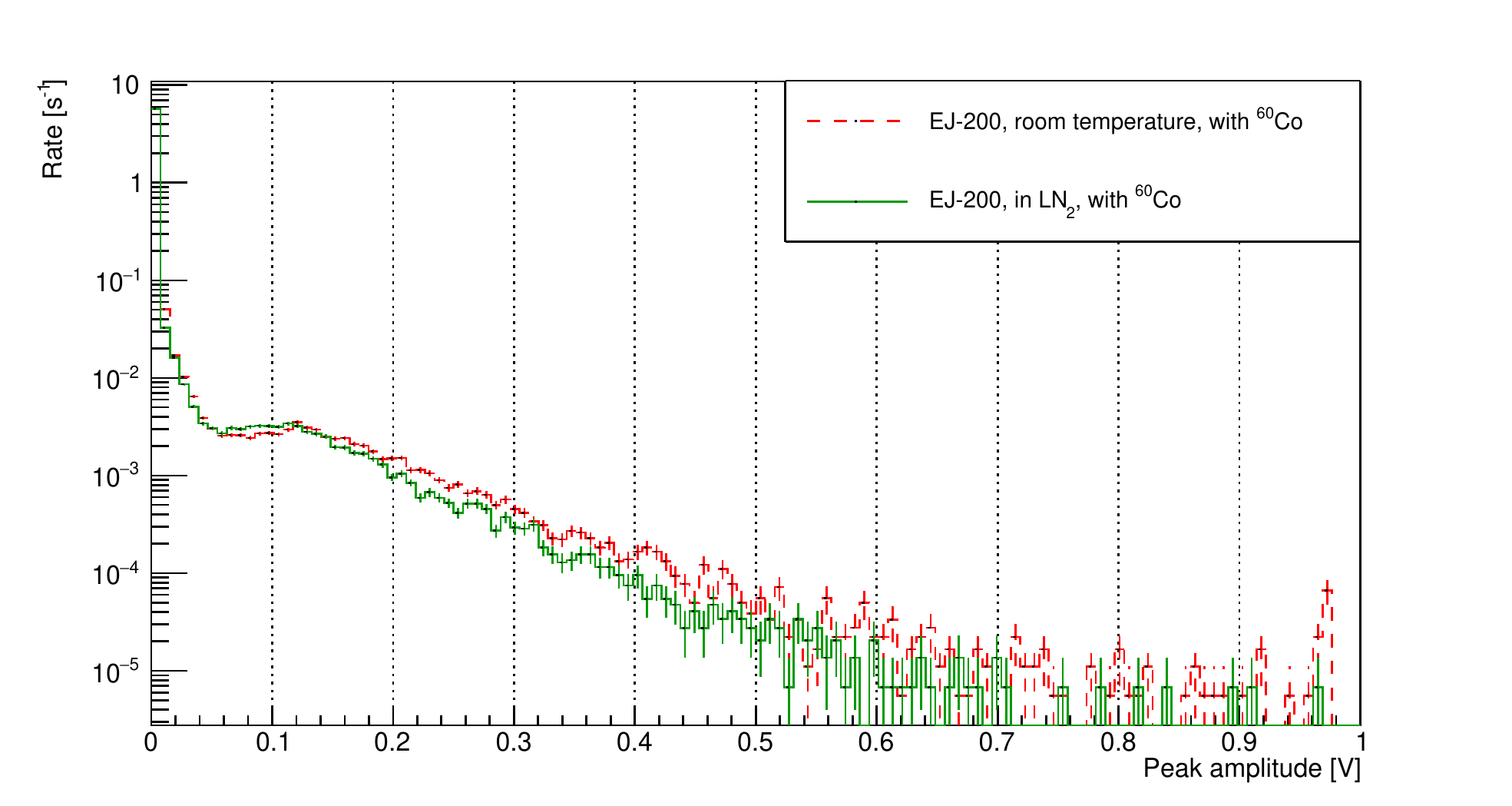}
    \caption{Typical trend of a scintillator's LY at room temperature and in LN$_2$. A sort of compression happens between the two curves.}
    \label{fig:Cryo_compression}
\end{figure}

\begin{figure}[!h]
    \centering
    \includegraphics[width=\textwidth]{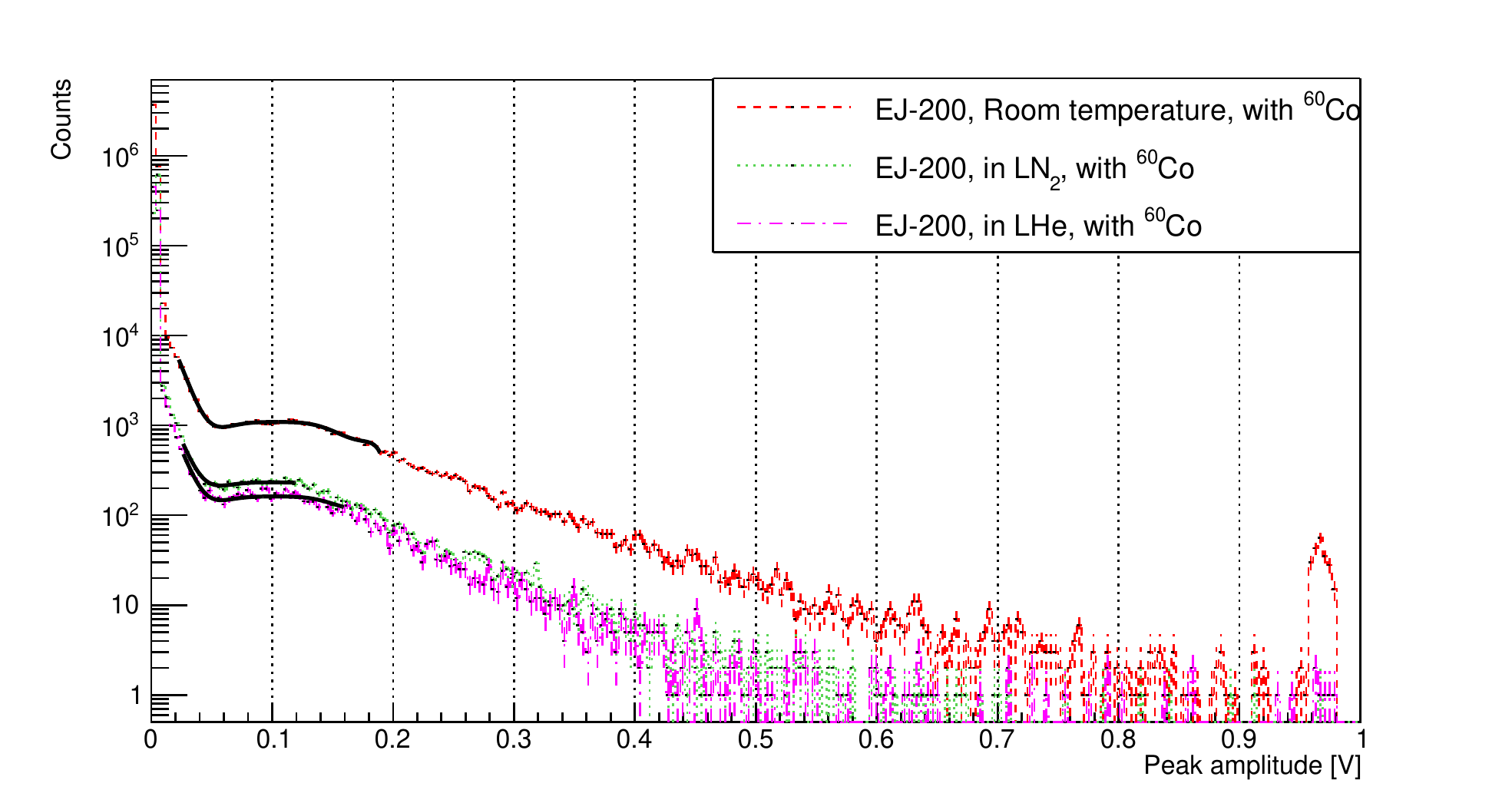}
    \caption{An example of the analysis performed. In this case, the curves were not normalized with the run duration. The black curves represent the fit performed on each spectrum, in different ranges which were chosen following a study to identify the most optimal range for each fit.}
    \label{fig:EJ200_fit}
\end{figure}

\begin{figure}[!h]
    \centering
    \includegraphics[width=\textwidth]{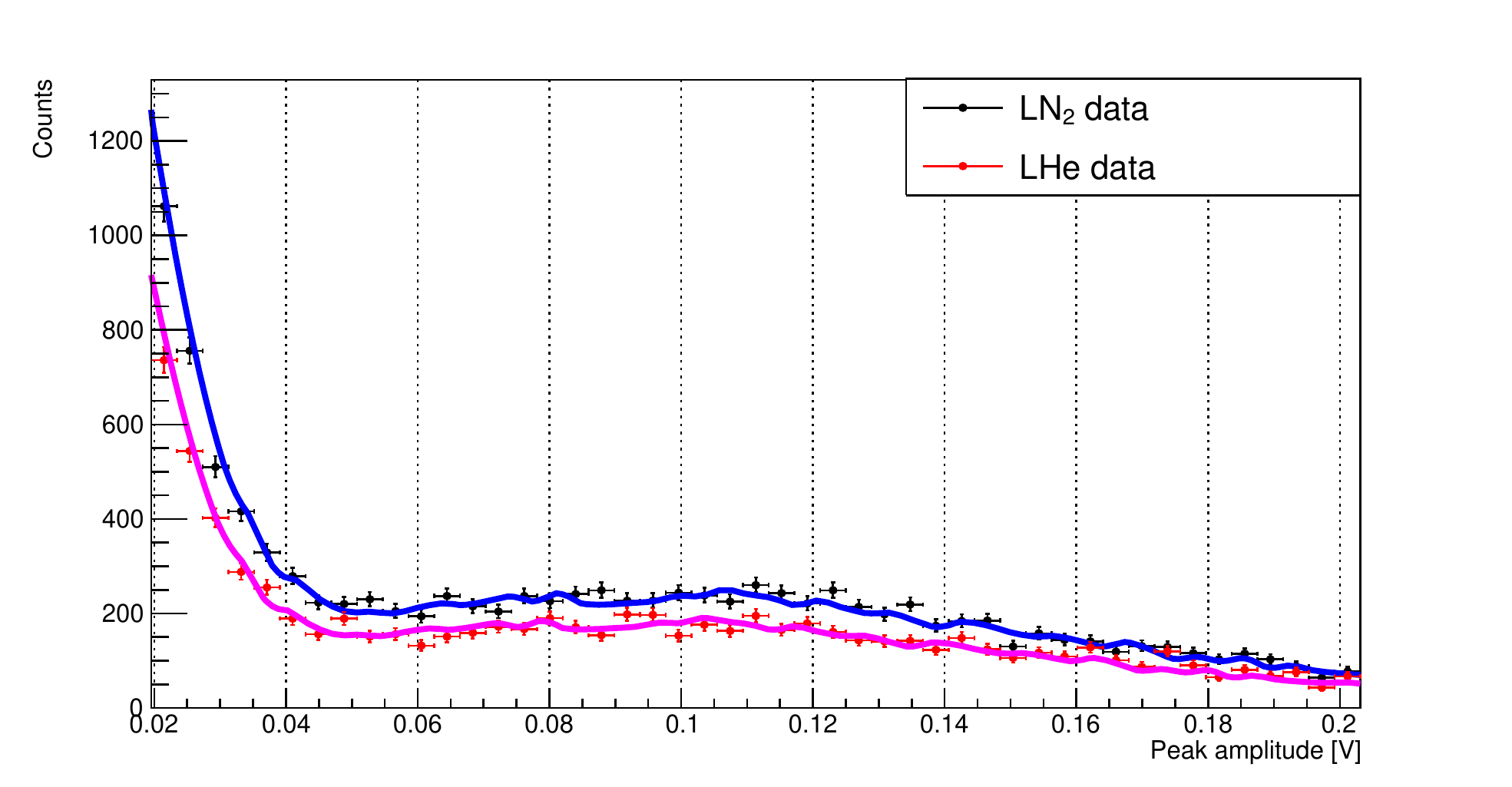}
    \caption{Trend of data acquired in LN$_2$ (black markers) and LHe (red markers) and relative fits performed with ROOFIT.}
    \label{fig:ROOFIT}
\end{figure}

The goal of the data analysis was to quantify the LY variation between ambient and cryogenic temperatures for each scintillator sample. Looking at the spectra reported in fig.~\ref{fig:LN2} and fig.~\ref{fig:LHe}, we can notice that at cryogenic temperatures the spectra of all scintillators look compressed with respect to the corresponding spectra at room temperature (see fig \ref{fig:Cryo_compression}).
The data analysis aims to quantify this compression factor (called $\alpha$) for each scintillator, as a proxy variable for the LY.
We modeled the spectral form due to the radioactive source, which allowed to have a higher statistic of low energy events, and the cosmic muons with an eight-degree polynomial, as reported in eq.~\ref{PESCE_fit_amb}:
\begin{equation}\label{PESCE_fit_amb}
    \sum_{k=0}^{8} p_kx^k
\end{equation}
The room temperature spectrum is then fitted with the polynomial, with the $p_k$ coefficient left as free parameters of the fit. 
The spectral form at cryogenic temperatures is modeled with the same polynomial (including the $p_k$ values fixed by the previous fit) except for two scaling factors as free parameters, one on the vertical axis $N$ (to account for the different number of events) and one on the horizontal axis ($\alpha$), which represents the relative change in LY, according to eq.~\ref{PESCE_fit_cryo}:
\begin{equation}\label{PESCE_fit_cryo}
   N \sum_{k=0}^{8} p_k\left(\frac{x}{\alpha}\right)^k
\end{equation}
The fit range is chosen by varying both the lower limit (between 0.01 and 0.03 V, with steps of 0.001 V) and the upper limit (between 0.1 and 0.3 V, with steps of 0.01 V), and selecting the fit with the smallest reduced $\chi^{2}$. 
The distribution of the results of the less performant fits is used to estimate the systematic error on the $\alpha$ parameter.
An example of this analysis, performed on the EJ-200 scintillator, is reported in fig.~\ref{fig:EJ200_fit}.
The $\alpha$ factors for the liquid nitrogen ($\alpha_{LN_{2}}$) coolings were evaluated for all the scintillators, while the factors for liquid helium ($\alpha_{LHe}$) were evaluated for the three scintillators (EJ-200, EJ-244, and EJ-248) that we evaluated most promising towards future large scale applications, i.e. those that had small changes in LY in LN$_2$.
The results are reported in tab.~\ref{tab:PESCE_alphas}.
\begin{table}
\centering
\begin{tabular}{ccc}
\hline
Scintillator & $\alpha_{LN_2}$ & $\alpha_{LHe}$ \\
\hline \hline
EJ-200 & 0.90 $\pm$ 0.01 $\pm$ 0.05  & 0.90 $\pm$ 0.01 $\pm$ 0.05 \\
EJ-244 & 0.89 $\pm$ 0.01 $\pm$ 0.01 & 0.93 $\pm$ 0.01 $\pm$ 0.03 \\
EJ-248 & 0.89 $\pm$ 0.01 $\pm$ 0.01 & 0.91 $\pm$ 0.01 $\pm$ 0.01 \\
EJ-208 & 0.83 $\pm$ 0.01 $\pm$ 0.02 & - \\
EJ-230 & 0.81 $\pm$ 0.01 $\pm$ 0.04 & - \\
EJ-240 & 0.90 $\pm$ 0.01 $\pm$ 0.02 & - \\
\hline
\end{tabular}
\caption{Evalutation of the $\alpha_{LN_2}$ and $\alpha_{LHe}$ factors for the tested commercial scintillators.}
\label{tab:PESCE_alphas}
\end{table}
We can see from tab.~\ref{tab:PESCE_alphas} that the $\alpha$ factors for LN$_2$ and LHe are substantially identical, meaning that there is no significant change in LY.
The scintillator which seems the most promising is EJ-200. For what concerns the scintillator EJ-240, a high $\alpha_{LN_{2}}$ is found: this happens because this device has a low LY at room temperature, which does not undergo large decreases at cryogenic temperatures. \\ 
For a further contribution to the systematics, the analysis was repeated for the scintillator EJ-200 using a template fit, i.e. ROOFIT~\cite{ROOFIT}, which is a toolkit for data modeling. ROOFIT provides a model distribution of observable x in terms of
parameters p, and derives a probability density function expressed as F(x,p).
Ranges of fits identical to those used for the polynomial fit were chosen. A model was built on the basis of the hot data, after which the histograms acquired at cold temperatures were fitted in the selected ranges. The results are illustrated in fig.~\ref{fig:ROOFIT}.
The fits were repeated for both distributions, varying the data intervals. The results obtained are reported in tab.~\ref{tab:alphas_roofit}.
\begin{table}[!h]
    \centering
\begin{tabular}{ccc}
\hline
Scintillator & $\alpha_{LN_2}$ & $\alpha_{LHe}$ \\
\hline \hline
EJ-200 &  0.92$\pm$0.1 &  0.93$\pm$0.04 \\
\hline
    \end{tabular}
    \caption{$\alpha$ values for scintillator EJ-200 obtained with ROOFIT.  }
    \label{tab:alphas_roofit}
\end{table}
As it can be observed, these results are comparable with those reported in tab.~\ref{tab:PESCE_alphas}. \\
In conclusion, it was possible to characterize these scintillators in a cryogenic environment, obtaining a satisfactory LY even at low temperatures and the efficiency comparisons of the scintillators are consistent with what was expected. \\

\section{Discussion}
The scintillator EJ-200 is the most performing: its LY does not seem to be affected by the temperature decrease, as can be seen from the cooling plots in fig.~\ref{fig:LHe} and from its $\alpha_{LN_2}$ and $\alpha_{LHe}$ values. In fact, as seen from the spectra, the curves at room temperature, in LN$_2$ and LHe almost superimposable. 
The scintillator EJ-208 shows instead a clear decrease in the LY when the system is at temperatures close to that of liquid nitrogen (fig.~\ref{fig:LN2}, tab.~\ref{tab:PESCE_alphas}). \\
For what concerns the scintillator EJ-230, no significant differences are noted in the LY at different temperatures, but the $\alpha_{LN_2}$ value is quite low (compared with the one of a good candidate such as EJ-200). Moreover, as reported in tab.~\ref{tab:PESCE_Eljen_scintillators}, this is the worst sample in terms of nominal light attenuation length. \\
The scintillator EJ-240 has the worst behavior: even though the $\alpha_{LN_2}$ is equal to the one of the scintillator EJ-200, from fig.~\ref{fig:LN2} can be seen that the effect of the $^{60}$Co is slightly noticeable (see blue and red dashed curves). On the other hand, this scintillator has the lowest concentration of scintillating molecule, as reported in tab.~\ref{tab:PESCE_Eljen_scintillators}, explaining this behavior. \\
As for the scintillator EJ-244, there is a reduction in terms of LY between the spectrum at room temperature and the spectrum acquired at cryogenic temperature (curves dashed red and dotted green in fig.~\ref{fig:LHe}), but, unlike the previous case, the effect of the source is well visible. Its $\alpha_{LN_2}$ and $\alpha_{LHe}$ values (tab.~\ref{tab:PESCE_alphas}) are also very similar to the ones of the best candidate, EJ-200.  Moreover, the light attenuation length (tab.~\ref{tab:PESCE_Eljen_scintillators}) is quite high (second only to the EJ-200 scintillator).
At last, the scintillator EJ-248 shows just a small decrease in LY in the region between 0.02 Volt and 0.04 Volt (fig.~\ref{fig:LHe}). Since it is a very small reduction (also confirmed by its $\alpha_{LN_2}$ and $\alpha_{LHe}$ values in tab.~\ref{tab:PESCE_alphas}), this device is still considered one of the best candidates.

\section{Conclusions}
Thanks to an adequate thermal coupling system, it was possible to successfully complete LY measurements of some commercial plastic scintillators (EJ-200, EJ-208, EJ-230, EJ-240, EJ-244, and EJ-248, by Eljen Technology) at room temperature, in liquid nitrogen, and in liquid helium, using a $^{60}$Co $\gamma$-ray source. 
The results show that the light yield of the commercial plastic scintillators does not undergo major variations at cryogenic temperatures. \\
Therefore, the possibility of building veto systems for rare event experiments with similar devices is realistic but, in this case, particular care is required in conforming to the stringent limits of radiopurity that these experiments require. \\
However, the results of this study are encouraging for future applications in rare event searches conducted in a cryogenic environment.


\acknowledgments
This project was completed thanks to the funds of the INFN 19593 grant, thanks to which it was also possible to finance the research contract of Dr. S. Caprioli. We also thank Dr. E. Ferri for her collaboration, the INFN computing center, and the INFN workshop and mechanical design service. Finally, we thank M. Rossi for his help in the development of the slow monitoring system. 



\bibliographystyle{aichej} 
\bibliography{bibliography.bib}

\begin{thebibliography}{10}
\providecommand{\url}[1]{\texttt{#1}}
\providecommand{\urlprefix}{URL }

\bibitem{Plastic_scint}
Łukasz Kapłon, Moskal G.
\newblock Blue-emitting polystyrene scintillators for plastic scintillation
  dosimetry.
\newblock \emph{Bio-Algorithms and Med-Systems}.
  2021;\hspace{0pt}17(3):191--197.

\bibitem{Berdugo:2022zzo}
Berdugo J.
\newblock {Latest Results of the Alpha Magnetic Spectrometer on the
  International Space Station}.
\newblock \emph{Moscow Univ Phys Bull}. 2022;\hspace{0pt}77(2):71--82.

\bibitem{ToF}
Lin WJ, Jianwei Z, Sun B, He LC, Lin W, Liu CY, Tanihata I, Terashima S, Tian
  Y, Wang F, Wang M, Zhang G, Zhang x, Zhu LH, Duan LM, Hu RJ, Liu Z, Lu CG,
  Ren PP, Zheng Y.
\newblock Plastic scintillation detectors for precision Time-of-Flight
  measurements of relativistic heavy ions.
\newblock \emph{Chinese Physics C}. 2016;\hspace{0pt}41.

\bibitem{ALEKSA2013442}
Aleksa M, Diemoz M.
\newblock Discussion on the electromagnetic calorimeters of ATLAS and CMS.
\newblock \emph{Nuclear Instruments and Methods in Physics Research Section A:
  Accelerators, Spectrometers, Detectors and Associated Equipment}.
  2013;\hspace{0pt}732:442--450.
\newblock Vienna Conference on Instrumentation 2013.

\bibitem{Gerda}
Agostini M, Bakalyarov A, Balata M, Barabanov I, Baudis L, Bauer C, Bellotti E,
  Belogurov S, Belyaev S, Benato G, Bettini A, Bezrukov L, Bode T, Borowicz D,
  Brudanin V, Brugnera R, Caldwell A, Cattadori CM, Chernogorov A, Zuzel G.
\newblock Upgrade for Phase II of the Gerda experiment.
\newblock \emph{The European Physical Journal C}. 2017;\hspace{0pt}78.

\bibitem{Legend_design}
Schwarz M, Krause P, Leonhardt A, Papp L, Schönert S, Wiesinger C, Fomina M,
  Gusev K, Rumyantseva N, Shevchik E, Zinatulina D, Araujo G.
\newblock Liquid Argon Instrumentation and Monitoring in LEGEND-200.
\newblock \emph{EPJ Web of Conferences}. 2021;\hspace{0pt}253:11014.

\bibitem{NEMO}
\relax{SuperNEMO} Collaboration.
\newblock Calorimeter development for the SuperNEMO double beta decay
  experiment.
\newblock \emph{Nuclear Instruments and Methods in Physics Research Section A:
  Accelerators, Spectrometers, Detectors and Associated Equipment}.
  2017;\hspace{0pt}868:98--108.

\bibitem{refId0}
{Biassoni, M}, {Brofferio, C}, {Faverzani, M}, {Ferri, E}, {Ghislandi, S},
  {Milana, S}, {Nutini, I}, {Pettinacci, V}, {Pozzi, S}, {Quitadamo, S}.
\newblock An acrylic assembly for low-temperature detectors.
\newblock \emph{Eur Phys J Plus}. 2021;\hspace{0pt}136(10):986.
\newline\urlprefix\url{https://doi.org/10.1140/epjp/s13360-021-01978-9}

\bibitem{Eljen}
Eljen Technology. 2021.
\newline\urlprefix\url{https://eljentechnology.com}

\bibitem{60Co}
Browne E, Tuli J.
\newblock Nuclear Data Sheets for {A = 60}.
\newblock \emph{Nuclear Data Sheets}. 2013;\hspace{0pt}114(12):1849--2022.

\bibitem{Labview}
LabVIEW, by National Instruments.
\newline\urlprefix\url{https://www.ni.com/it-it/shop/labview.html}

\bibitem{ROOT}
ROOT Data Analysis Framework.
\newline\urlprefix\url{https://root.cern/about/}

\bibitem{ROOFIT}
W~Verkerke DK.
\newblock RooFit Users Manual.
\newline\urlprefix\url{https://root.cern.ch/download/doc/RooFit_Users_Manual_2.91-33.pdf}

\end{thebibliography}



\end{document}